\RequirePackage[undo-recent-deprecations]{expl3}
\documentclass[a4paper,fleqn]{cas-sc}
\usepackage{soul}
\usepackage[sort&compress,numbers]{natbib}
\usepackage{lineno}
\usepackage{setspace}
\usepackage{verbatim}
\usepackage{graphics}
\usepackage{graphicx}
\usepackage{cuted}
\usepackage{lipsum}
\usepackage{multirow}

\def\tsc#1{\csdef{#1}{\textsc{\lowercase{#1}}\xspace}}
\tsc{WGM}
\tsc{QE}
\tsc{EP}
\tsc{PMS}
\tsc{BEC}
\tsc{DE}

\begin{document}

\let\WriteBookmarks\relax
\def\floatpagepagefraction{1}
\def\textpagefraction{.001}
\shorttitle{Adsorption Mechanism of Caffeine on MAPbI$_3$ Perovskite Surfaces}
\shortauthors{Pereira J\'unior \textit{et~al}.}

\title [mode = title]{On the Adsorption Mechanism of Caffeine on MAPbI$_3$ Perovskite Surfaces: A Combined UMC--DFT Study}

\author[1,2]{Luiz A. Ribeiro J\'unior}
\cormark[1]
\ead{ribeirojr@unb.br}
\author[3]{Raphael M. Tromer}
\author[1]{Ramiro M. dos Santos}
\author[3,4]{Douglas S. Galv\~ao}


\address[1]{Institute of Physics, University of Bras\'ilia, 70910-900, Bras\'ilia, Brazil}
\address[2]{PPGCIMA, Campus Planaltina, University of Bras\'{i}lia, 73345-010, Bras\'{i}lia, Brazil}
\address[3]{Applied Physics Department, University of Campinas, Campinas, S\~ao Paulo, Brazil.}
\address[4]{Center for Computing in Engineering and Sciences, University of Campinas, Campinas, S\~ao Paulo, Brazil.}

\begin{abstract}
Recently, it was experimentally shown that the performance and thermal stability of the perovskite MAPbI$_3$ were improved upon the adsorption of a molecular layer of caffeine. In this work, we used a hybrid methodology that combines Uncoupled Monte Carlo (UMC) and Density Functional Theory (DFT) simulations to carry out a detailed and comprehensive study of the adsorption mechanism of a caffeine molecule on the surface of MAPbI$_3$. Our results showed that the adsorption distance and energy of a caffeine molecule on the MAPbI$_3$ surface are 2.0 \r{A} and -0.3 eV, respectively. The caffeine/MAPbI$_3$ complex presents a direct bandgap of 2.38 eV with two flat intragap bands distanced 1.15 and 2.18 eV from the top of valence bands. Although the energy band levels are not significantly shifted by the presence of caffeine, the interaction MAPbI$_3$/perovskite is enough to affect the bands' dispersion, particularly the conduction bands.
\end{abstract}


\maketitle
\doublespacing

\section{Introduction}
The achievements obtained in producing a more efficient third-generation of solar cells were fundamental for the recent advances in photovoltaics \cite{green2002third,nozik2010semiconductor,conibeer2006silicon,ludin2014review,polman2012photonic}. Nowadays, this new generation of photovoltaic technology, with low cost and potential fast production on a large scale, has as one of the protagonists the organic-inorganic halide perovskites \cite{green2014emergence,liu2013efficient,zhou2014interface,snaith2014anomalous,correa2017promises,tsai2016high}. Among these perovskites, the most studied material is MAPbI$_3$ (where MA refers to methylamine) \cite{wu2017thermally,chen2019single,li2018improved,jiang2016post,abdelmageed2016mechanisms,frolova2015chemical,you2018biopolymer,yuan2015photovoltaic,heo2015hysteresis,han2015degradation,cha2016enhancing,wang2019caffeine}. The success of this material in manufacturing solar cells is attributed to some of its outstanding properties such as suitable bandgap for optoelectronics applications, high photon absorption coefficient, fast exciton to polaron conversion (due to very low biding energy for excitons), and the long lifetime presented by free charge carriers \cite{green2014emergence,correa2017promises}. Although MAPbI$_3$ possesses interesting properties for photovoltaic applications, the problem of low photo and thermal stability are still drawbacks for its large-scale use \cite{correa2017promises,wu2017thermally,han2015degradation}.   

Recently, experimental investigations proposed improvements in the thermal stability of perovskite-based solar cells (PSCs) upon the doping or adsorption of small molecules (molecular dyes) on their surfaces \cite{ludin2014review,wang2019caffeine,bella2016improving,choi2018thermally,park2015perovskite,yun2018new,wang2016triarylamine,xu2014carbazole,pathak2014performance,xu2016lowcost}. One of the aims of these studies was to address the fast degradation of these perovskite species when exposed to environmental factors. Wang and colleagues showed that the adsorption of caffeine molecules improved the performance and thermal stability of PSCs based on MAPbI$_3$ layers \cite{wang2019caffeine,wang2019constructive}. Their findings revealed that the strong interaction between caffeine and Pb$^{2+}$ ions serves as a ``molecular lock'' that increases the activation energy during film crystallization, yielding a perovskite film with a preferred orientation, improved electronic properties, greatly enhanced thermal stability, and a stabilized power conversion efficiency (PCE) of 19.8\% \cite{wang2019caffeine}. Albeit some dye-sensitized PSCs have been proposed recently with improved PCE \cite{park2015perovskite,pathak2014performance,xu2014carbazole}, the adsorption mechanisms between molecular dyes on perovskite layers remain not fully understood.

In the present work --- inspired by the recent progress in obtaining MAPbI$_3$-based PSCs with improved performance upon caffeine doping \cite{wang2019caffeine,wang2019constructive} --- we carried out uncoupled Monte Carlo (UMC) and density functional theory (DFT) simulations to study the structural and electronic properties of the caffeine/MAPbI$_3$ complex. Caffeine and MA (methylammonium) have the following chemical formulas C$_8$H$_{10}$N$_4$O$_2$ and CH$_3$NH$_3$, respectively. This hybrid methodology showed a good compromise between computational cost and accuracy \cite{leal2017combined}. From the UMC realizations, the caffeine/MAPbI$_3$ conformation of the lowest interaction energy is obtained. DFT calculations were used to obtain the electronic properties of this structure. In this way, by combining UMC and DFT methodologies we proposed a more detailed and comprehensive description of the adsorption mechanism of caffeine molecules on MAPbI$_3$ layers \cite{wang2019constructive}, contributing to further understand the interaction between these species.          

\begin{figure}[pos=ht]
\centering
\includegraphics[width=\linewidth]{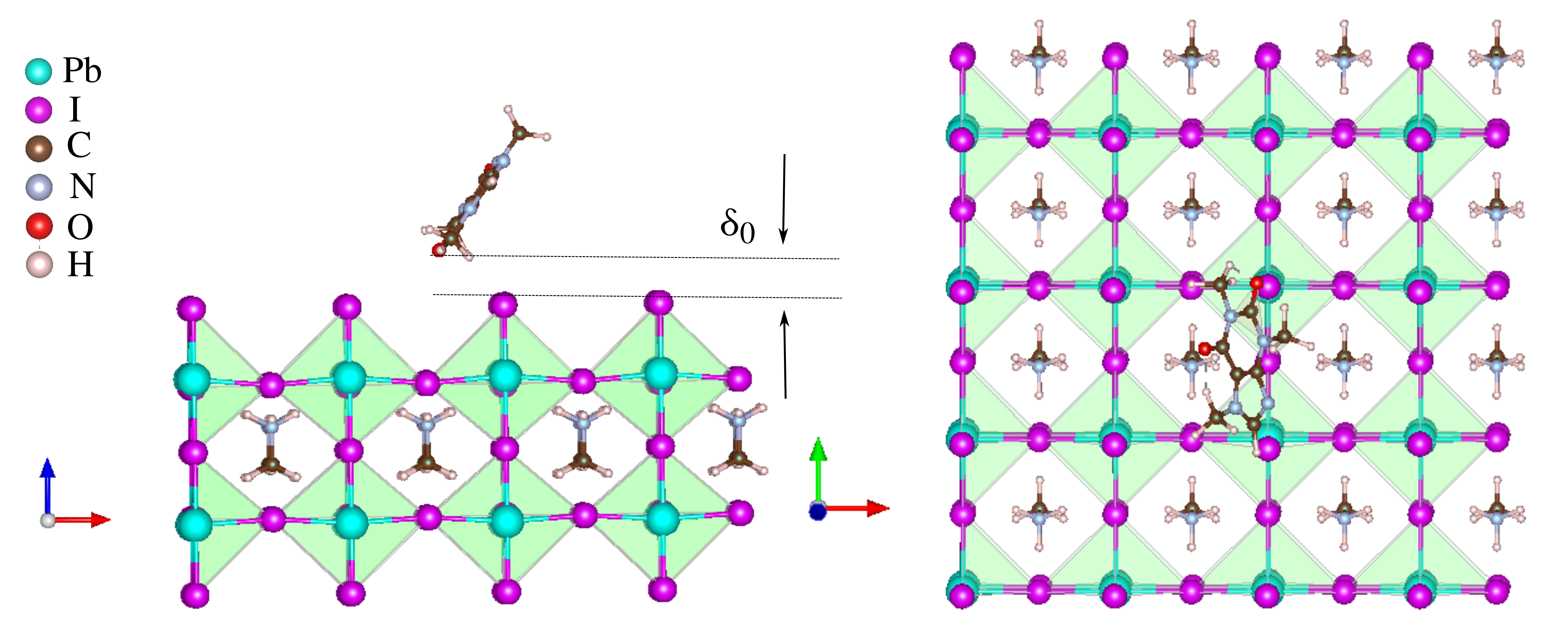}
\caption{Schematic representation of the DFT optimized caffeine/MAPbI$_3$ structure. The left panel illustrates the side view whereas the right panel shows the top view. The color scheme for the atoms is presented within the figure. Caffeine and MA (methylammonium) have the following chemical formulas C$_8$H$_10$N$_4$O$_2$ and CH$_3$NH$_3$, respectively. $\delta_0$ is the distance between this molecule to the perovskite surface after the geometry optimization (2.0 \r{A}).}
\label{fig:system}
\end{figure} 

\section{Methodology}

For the description of the structural and electronic properties of a caffeine molecule adsorbed on a MAPbI$_3$ layer, we used UMC (Adsorption Locator Modulus) as implemented in Materials Studio  \cite{vcerny1985thermodynamical,frenkel2002understanding,Kirkpatrick671,doi:10.1063/1.1699114}. The DFT calculations were carried out using the SIESTA code \cite{Soler_2002}. The UMC approach samples several configurations in the canonical ensemble by generating a large set of caffeine/MAPbI$_3$ systems, starting from a trial (random) configuration. The arrangement of these conformations is randomly chosen by rotating and translating the molecule (caffeine) about an axis fixed by a substrate (MAPbI$_3$ layer). The search for the lowest energy configurations is obtained by using the simulated annealing method. The Metropolis algorithm provides the statistical weights employed in this scheme \cite{doi:10.1063/1.1699114}. Here, the parameters used in the annealing procedure were: 50000 annealing cycles, 300 K for the initial temperature, 500 K for the mid-cycle temperature, 5 heating ramps per cycle, 100 dynamic steps per ramp, and 5000 for the total number of steps. These parameters were used before to simulate the interaction between a pentacene dimer \cite{leal2017combined}. Importantly, a satisfactory number of conformations (100) were considered. The UMC mentioned above was repeated 100 times to obtain the best set of minimum-energy configurations. This UMC protocol yields the system conformation with the lowest interaction energy (see Figure \ref{fig:system}). There are two other possible configurations, with MA or Pb ions covering (see Figures S1 and S2(a) in the Supplementary Material). The configuration in \ref{fig:system} is the most stable one \cite{wang2019caffeine,wang2019constructive}. We take these complexes to study the electronic properties of the caffeine molecule adsorbed on MAPbI$_3$.  

When it comes to the DFT calculations, a good simulation performance in describing a system containing the Pb$^{2+}$ ion is achieved using single zeta polarized basis functions \cite{PhysRevB.64.195103}. Relativistic pseudopotentials were also used \cite{PhysRevB.43.8861}, with core correction and parameterized by Troullier-Martins' method \cite{PICKETT1989115}. The exchange-correlation energies were approximated by the GGA functional, using Becke's exchange functional methodology \cite{PhysRevA.38.3098}, and Lee-Yang-Parr correlation (BLYP) \cite{PhysRevB.37.785}. The BLYP functional was successfully used to calculate the electronic structure properties of perovskites \cite{BARHOUMI2018120,Yine1701793}. The cut-off energy was settled as 500 Ry. Due to the large value of the lattice parameters, we have employed a $15\times 15\times 5$ Monkhorst-Pack mesh to better describe the properties within a volume in the reciprocal lattice \cite{PhysRevB.13.5188}. Importantly, the lowest energy configuration yielded in UMC simulations was optimized using DFT and the final MA orientation was practically unaltered, while, as expected, the perovskite geometry is slightly different.

\begin{figure*}[pos=ht]
\centering
\includegraphics[width=0.8\linewidth]{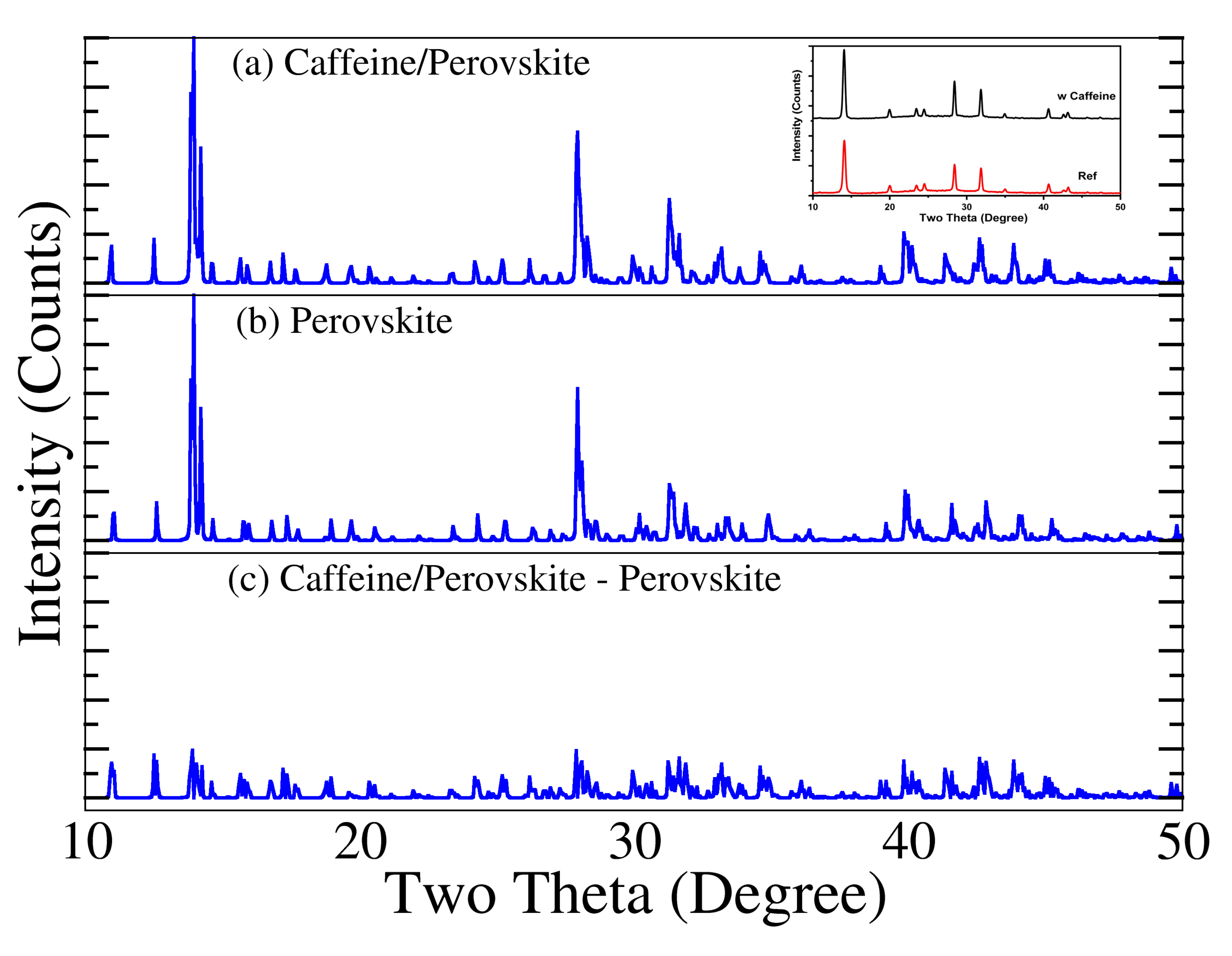}
\caption{Simulated X-ray diffraction pattern for the caffeine/MAPbI$_3$ configuration at 300 K, after the dye adsorption. (a) The complex perovskite/caffeine, (b) isolated perovskite, (c) the difference between the data shown in (a) and (b)}. The inset panel shows the XRD measurements obtained in reference \cite{wang2019caffeine}.
\label{fig:xray}
\end{figure*}

\section{Results}

We begin our discussions by examining the crystal structure for the caffeine/MAPbI$_3$ configuration after the dye adsorption. Figure \ref{fig:system} shows the DFT optimized geometry for the caffeine/MAPbI$_3$ complex. $\delta_0$ is the distance between this molecule to the perovskite surface (Iodine atoms) after the geometry optimization 2.0 \r{A}. The perovskite present cubic structure with supercell dimensions of 25.57 \r{A} $\times$ 25.57 \r{A} $\times$ 56.7 \r{A} for the x, y, and z directions, respectively. The value along the z-direction is large enough to prevent spurious interactions between the image layers. The Pb-I distances are 3.20 \r{A} and 3.22 \r{A} along the x and y directions, respectively, and 2.90 \r{A} for z-direction. The forces converged until reaching a minimum value of 1.0$\times 10^{-3}$ eV/\AA. To ensure a good compromise between the accuracy of our results and the computational feasibility, the tolerance in the matrix and the total energy was set at 1.0$\times 10^{-4}$ and 1.0$\times 10^{-5}$ eV, respectively.

\begin{figure}[pos=ht]
\centering
\includegraphics[width=0.8\linewidth]{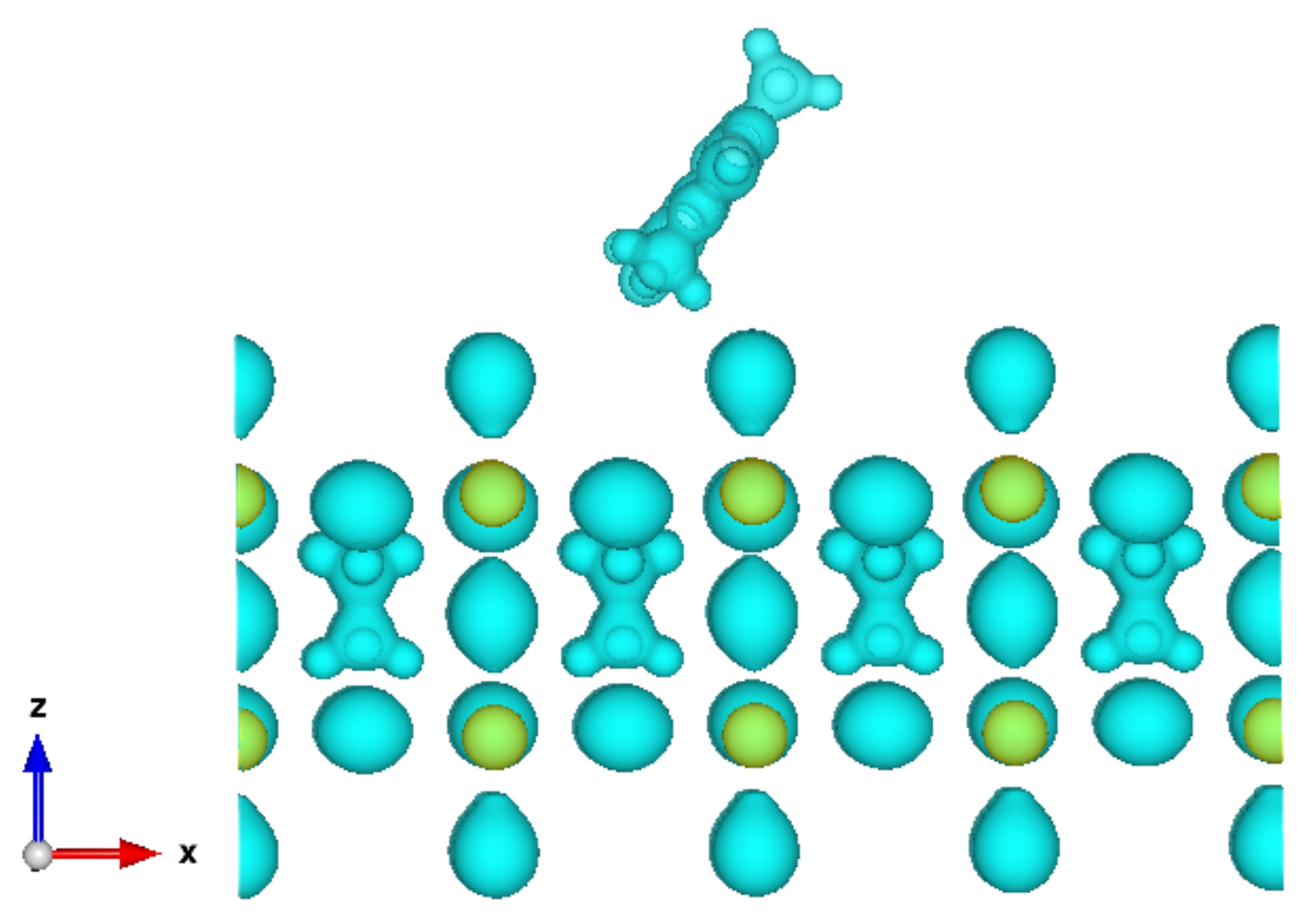}
\caption{Schematic representation of the electrostatic potential of caffeine/perovskite. Green and cyan represent negative and positive values, respectively.}
\label{fig:eletropot}
\end{figure}

Figure \ref{fig:xray}(a) depicts the simulated X-ray diffraction (XRD) pattern for this configuration at 300 K. Figures \ref{fig:xray}(b) and \ref{fig:xray}(c) show the XRD pattern for the isolated perovskite layer and for the difference between the cases \ref{fig:xray}(b) and \ref{fig:xray}(c), respectively. The XRD pattern was obtained by using the Reflex module as implemented in Materials Studio \cite{baur1992perils,berar1993modeling,Boultif:la0010}. These XRD patterns were simulated using the optimized structure presented in Figure \ref{fig:xray}. It was used as a benchmark that our results are consistent with the experimental one. In our calculations, we used a single crystal and the following diffractometer parameters: a two theta range of [5-45] degrees with a step size of 0.05 degrees; the distance between the planes of atoms that give rise to diffraction peaks (1/$d_{}$) were settled to 0.056628 and 0.49681 for the minimum and maximum theta two values, respectively; and the Bragg–Brentano geometry was employed. For the radiation, we used in the simulations a copper source, $\lambda_1$ = 1.540 \r{A}, $\lambda_2$ = 1.544 \r{A}, $I_2$/$I_1$ = 0.50 (the intensity ratio of the x-rays corresponding to the wavelengths $\lambda_2$ and $\lambda_1$, respectively), and polarization of 0.5 (that specifies the fraction of the x-ray beam polarized in the direction perpendicular to the plane of the incident and diffracted beams).

\begin{figure*}[pos=ht]
\centering
\includegraphics[width=\linewidth]{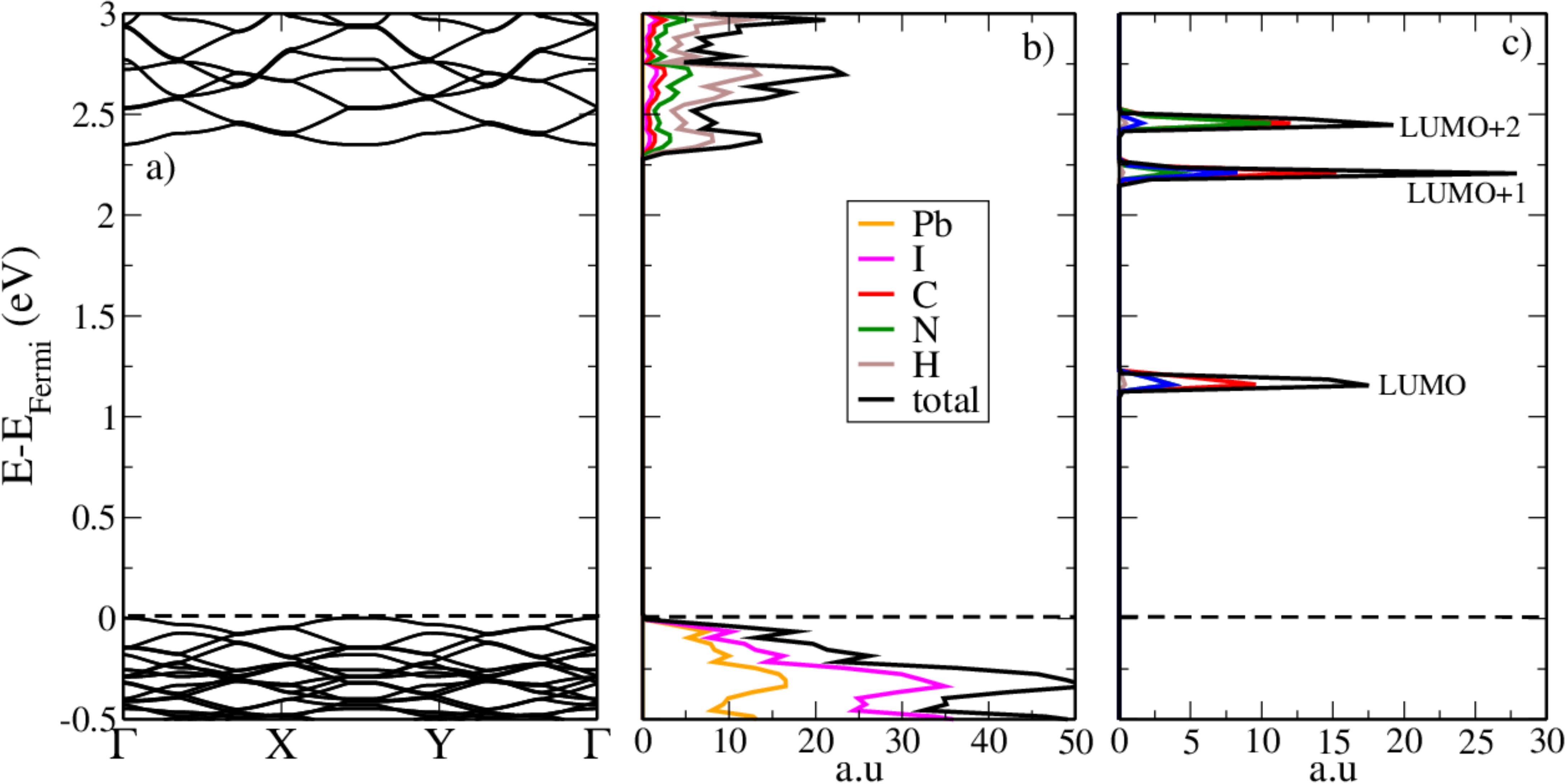}
\caption{(a) Electronic band structure and (b) the corresponding projected density of states (PDOS) for the isolated MAPbI$_3$ perovskite. (c) LUMO, LUMO+1, and LUMO+2 molecular orbitals for the isolated caffeine.}
\label{fig:bands}
\end{figure*}

\begin{figure*}[pos=ht]
\centering
\includegraphics[width=\linewidth]{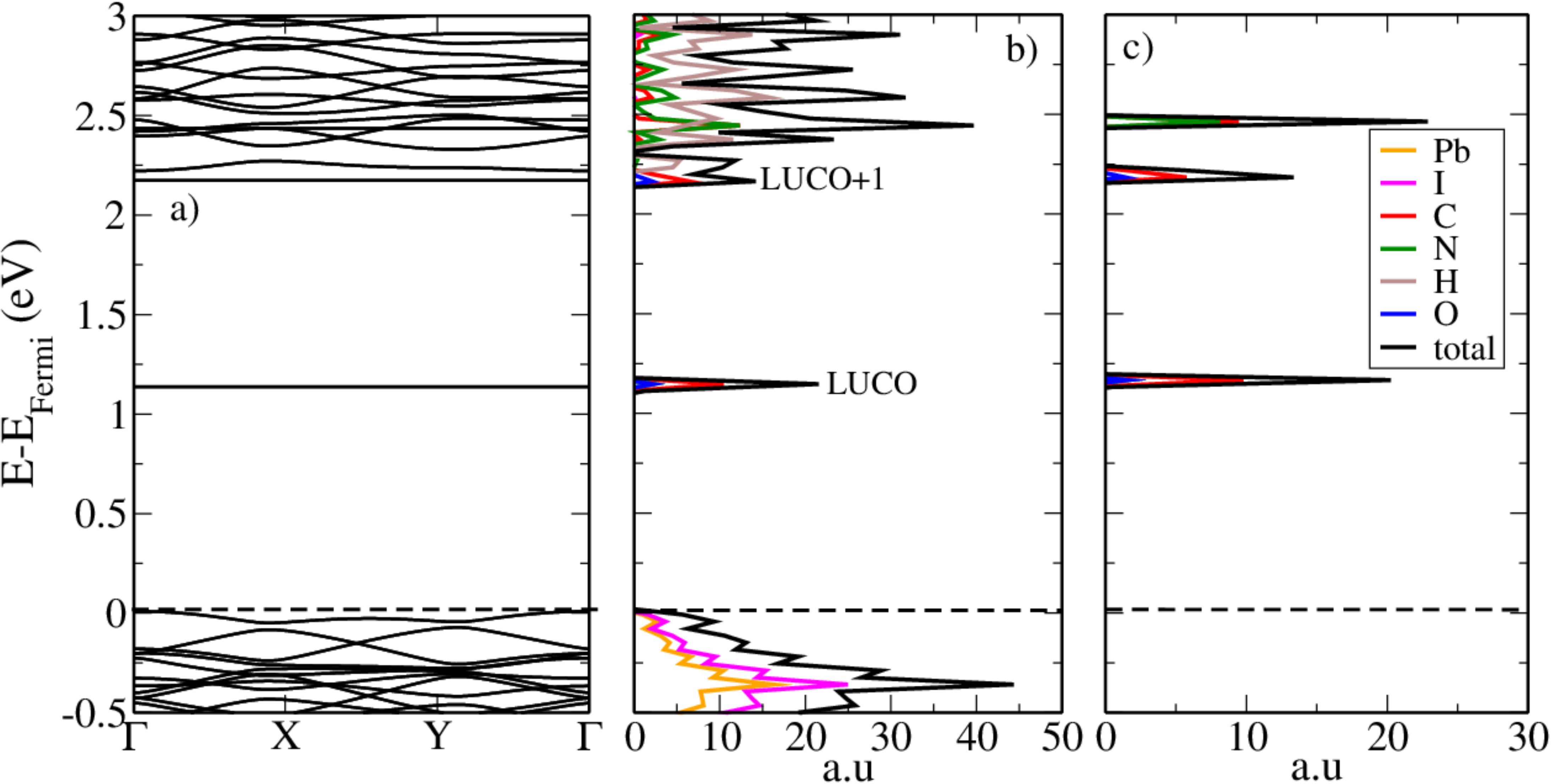}
\caption{(a) Electronic band structure and (b) the corresponding projected density of states (PDOS) for the MAPbI$_3$/caffeine system. (c) The contribution of the caffeine atoms to the PDOS.}
\label{fig:bands2}
\end{figure*}

The inset panel shows the XRD measurements obtained in reference \cite{wang2019caffeine}. We can see that the simulated XRD pattern shows a good agreement with the experimental one \cite{wang2019caffeine}, even considering the adsorption mechanism of a single caffeine molecule on top of a MAPbI$_3$ surface. The calculated reflection peaks occur at approximately 14.02$^{\circ}$, very close to the ones reported in the experiments (13.90$^{\circ}$, inset panel in Figure \ref{fig:xray}) \cite{wang2016triarylamine}. This XRD pattern points to a tetragonal perovskite phase with the dominant (110) lattice reflection at 14.02$^{\circ}$, which is the preferred orientation for the perovskite films \cite{wang2019caffeine}. These results show that our combined UMC--DFT approach can accurately describe the adsorption configuration of molecular dyes on the surface of MAPbI$_3$ layers. As expected, the caffeine adsorption does not significantly affect the main peaks associated with MAPbI$_3$ since caffeine remains on the surface (no intercalation), as shown in Figure \ref{fig:xray}(c). 

\begin{figure*}[pos=ht]
\centering
\includegraphics[width=\linewidth]{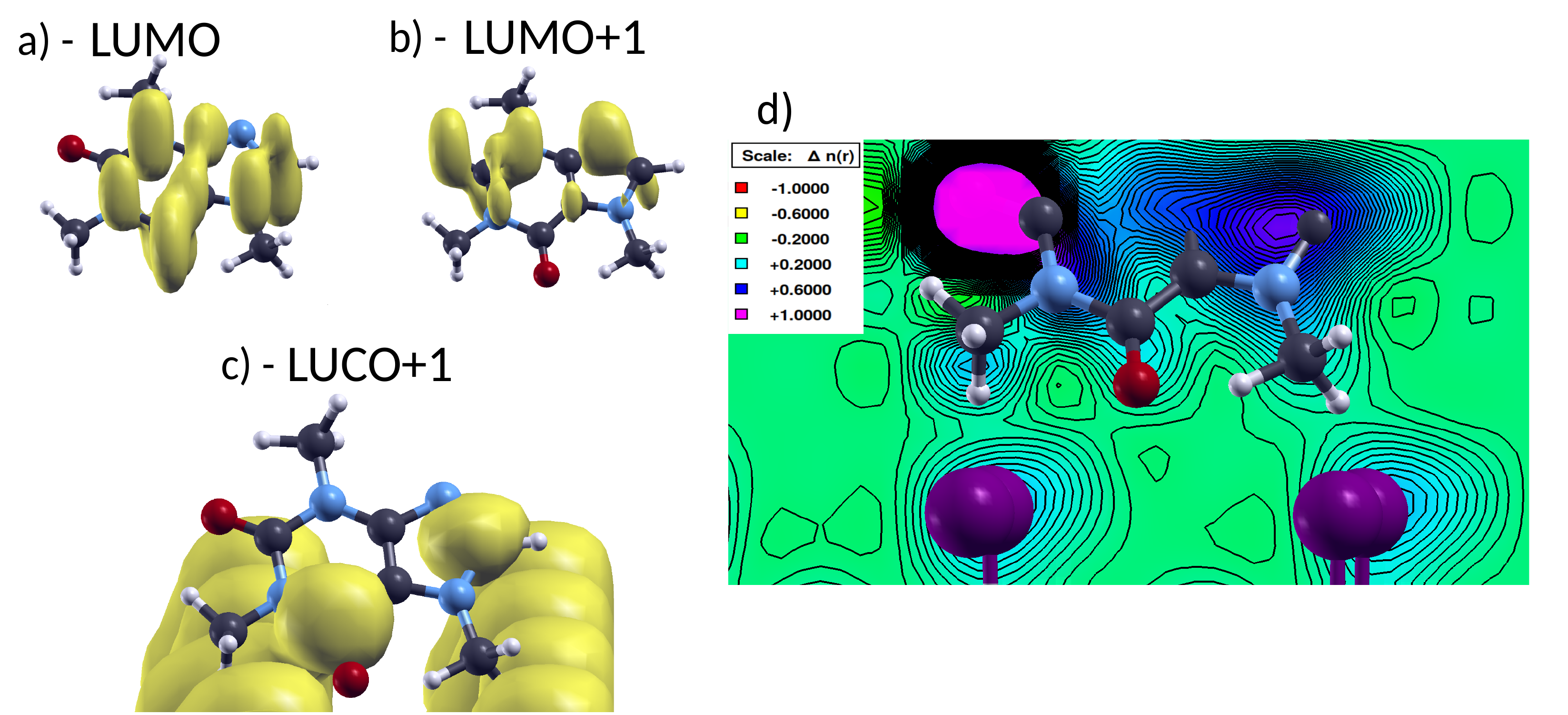}
\caption{(a) LUMO and (b) LUMO+1 (frontier orbitals) for an isolated caffeine molecule and, (c) LUCO+1 of the caffeine/MAPbI$_3$ complex. (d) depicts the isosurfaces charge density. The color code represents the values of the eleconic densities ($\delta n(r)$). The pink and red colors refer to the maximum ($+1e$, hole) and minimum ($-1e$, electron) electronic densities, respectively.}
\label{fig:ldos}
\end{figure*}

To quantify the interaction between caffeine and MAPbI$_3$, we have calculated the DFT adsorption energies accordingly to the expression $E_a=[E_{pc}-(E_{p}+E_c)]$, where $E_{pc}$,  $E_{p}$, and $E_c$ are the total energies for the caffeine/MAPbI$_3$ configuration after the dye adsorption, isolated MAPbI$_3$ monolayer, and isolated caffeine molecule, respectively. The DFT estimated adsorption energy is $E_a=-0.3$ eV.

Figure \ref{fig:eletropot} illustrates the electrostatic potential over the structure. Due to this electrostatic potential configuration, the caffeine molecule is adsorbed on the perovskite structure by interacting with I$^-$ and the Pb$^{2+}$ ions, which contributes to the stability of the structure. As claimed by Wang \textit{et al.} \cite{wang2019caffeine,wang2019constructive}, it is this strong Pb$^{2+}$/caffeine interaction that works as a  ``molecular lock'' enhancing the structural stability. It is important to stress that although the $E_a$ is larger for the case of MA covering (-1.14 eV), in this configuration the MA molecules can easily evaporate, which compromises the overall structural gain. For the Pb covering case, this configuration is structurally unstable (see Figure S2 (b) in the Supplemental Material).

Now we present the electronic band structure (Figure \ref{fig:bands} (a)) and the corresponding projected density of states (PDOS) (Figure \ref{fig:bands} (b)) for the isolated MAPbI$_3$ slab. In Figure \ref{fig:bands} (c), we show the LUMO, LUMO+1, and LUMO+2 states for the isolated caffeine molecule. LUMO refers to Lowest Unoccupied Molecular Orbital. In Figure \ref{fig:bands}(a), we can observe a direct bandgap of about 2.4 eV at the gamma point. This value agrees with the ones reported in the literature for other theoretical studies \cite{article,doi:10.1063/1.4930070}. It should be stressed that, in general, GGA tends to underestimate the bandgap values. However, there are a few cases in which GGA predicted band gaps in good agreement with experiments. For example, it was reported a calculated bandgap of about 1.6 eV for MAPbI$_3$ employing PBE-GGA \cite{gengATS} (probably due to error cancellations). There are also GGA studies where the perovskite bandgap values are overestimated \cite{gengATS}, as in our case. Here, we have also tested other MAPbI$_3$ slabs (see Supplemental Material). The obtained bandgap value for the stable configuration (MA covering) is about $1.4$ eV. We can see from the PDOS of the slab that I and Pb atoms have a large contribution to the valence bands, while the I and H atoms to the conduction ones. For the sake of comparison, we show in Figure \ref{fig:bands} (c) the LUMO, LUMO+1, and LUMO+2 molecular orbitals for the isolated caffeine.

In Figure \ref{fig:bands2} we present the electronic band structure \ref{fig:bands2}(a) and the corresponding PDOS \ref{fig:bands2}(b) of the caffeine/MAPbI$_3$ complex. Besides, we included in Figure \ref{fig:bands2}(c) the contributions of the caffeine atoms to the PDOS complex. Upon the caffeine adsorption, the MAPbI$_3$ bandgap is almost the same, with the appearance of two midgap flat bands (LUCO and LUCO+1, distanced 1.15 and 2.18 eV from the HOCO bands. HOCO and LUCO refer to the Highest/Lowest Occupied/Unoccupied Crystalline Orbitals, respectively. These flat bands can be attributed to caffeine and have almost the same energy values from the molecular case (Figure \ref{fig:bands}(c)). Since the momentum conservation of the excited electron upon photon absorption is characterized by the presence of a direct bandgap, the vertical (direct) transitions of this excited electron can form excitons spontaneously. The presence of flat midgap states favors the formation of excitons with considerably lower binding energy. These excitons are easily dissociated in the material due to the weak interaction between the electron-hole pair. We see from PDOS of caffeine/MAPbI$_3$ complex (Figure \ref{fig:bands2}(b)) that Pb and I atoms have large contributions to the valence bands, while the non-metal atoms have larger contributions to the conduction bands. As discussed below, there is a charge transfer of $-0.1e$ from caffeine to perovskite. Although this charge amount is not enough to significantly change the energy band positions, it is enough to affect the band's dispersion, as evidenced by contrasting Figures \ref{fig:bands}(a) and \ref{fig:bands2}(a).

Finally, Figure \ref{fig:ldos} shows (a) LUMO and (b) LUMO+1 (frontier orbitals) for an isolated caffeine molecule and, (c) LUCO+1 of the caffeine/MAPbI$_3$ complex. Figure \ref{fig:ldos}(d) depicts the isosurface charge density. The LUMO (isolated caffeine) and LUCO (complex, not shown in this figure) present similar orbital configurations (the LUMO is illustrated in Figure \ref{fig:ldos}(a)). On the other hand, the LUMO+1 (isolated caffeine) and LUCO+1 (complex) differ substantially when it comes to their distribution. Upon adsorption, as mentioned above, it occurs a charge transfer from caffeine to the perovskite. This charge transference is responsible for changing the orbital configurations of the related LUMO+1 and LUCO+1 levels. Figure \ref{fig:ldos}(d) shows a top view of the system presented in Figure \ref{fig:system} with its related isosurfaces. This particular view hides almost all the perovskite surface --- just 4 Iodine atoms (purple spheres) can be visualized --- and also some atoms of the caffeine molecule. The colors in Figure  \ref{fig:ldos}(d) denote the local variation in the electronic density ($\Delta$n(r)). In the adopted color code, the pink and red colors refer to the maximum (+1e, hole) and minimum (-1e, electron) electronic densities, respectively. The other colors stand for intermediate values for the electronic densities. One can note a continuous isosurface enveloping the hydrogen atoms (white spheres) and the Iodine ones (purple spheres), which shows the interaction channels between caffeine and perovskite for the model complex studied here.

\section{Conclusions}

In summary, we used a hybrid methodology that combined uncoupled Monte Carlo and density functional theory simulations to study the structural and electronic properties of a caffeine molecule adsorbed on the surface of MAPbI$_3$ perovskite. Our findings showed that the adsorption distance and adsorption energy of a caffeine molecule on the MAPbI$_3$ surface are 2.0 \r{A} and -0.3 eV, respectively. The calculated X-ray diffraction pattern at 300 K showed a good agreement with the experimental one. The calculated reflection peak occurs at approximately 14.02$^{\circ}$, very close to the one reported in experiments (13.90$^{\circ}$) \cite{wang2016triarylamine}. Moreover, the electrostatic potential analysis revealed that the caffeine molecule is adsorbed on the perovskite structure due to its strong interaction with the Pb$^{2+}$ ions, with a charge transfer of $0.1e$ from caffeine to perovskite. The caffeine/MAPbI$_3$ complex presents a direct bandgap of 2.38 eV with two flat intragap bands distanced 1.15 and 2.18 eV from the top of valence bands. Although the energy band levels are not significantly shifted by the presence of caffeine, the interaction MAPbI$_3$/perovskite is enough to affect the band's dispersion, particularly the conduction bands. Although the $E_a$ is larger for the case of MA covering (-1.14 eV), in this configuration the MA molecules can easily evaporate, which compromises the overall structural gain. For the Pb covering case, this configuration is structurally unstable. Pb and I atoms have large contributions to the valence bands, while the non-metal atoms have larger contributions to the conduction bands. We found a charge transfer of $-0.1e$ from caffeine to perovskite. Although this charge amount was not enough to significantly change the energy band positions, it was enough to affect the band structure dispersion. The improvements on the perovskite quality samples due to the caffeine presence were already unambiguously experimentally established \cite{wang2019caffeine,wang2019constructive}. Our results from \textit{ab initio} calculations and Monte Carlo techniques provide deeper insights into the structural and electronic aspects of the caffeine presence. We hope the present work can stimulate further studies on other dyes.

\section*{Acknowledgments}
The authors gratefully acknowledge the financial support from Brazilian Research Councils CNPq, CAPES, and FAPDF and CENAPAD-SP for providing the computational facilities. L.A.R.J. gratefully acknowledges respectively, the financial support from FAP-DF  and CNPq grants $00193.0000248/2019-32$ and $302236/2018-0$. D.S.G. thank the Center for Computing in Engineering and Sciences at Unicamp for financial support through the FAPESP/CEPID Grants $\#2013/08293-7$ and $\#2018/11352-7$. L.A.R.J. gratefully acknowledges the financial support from DPI/DIRPE (Edital $03/2020$) grant $23106.057541/2020-89$ and from IFD/UnB (Edital $01/2020$) grant $23106.090790/2020-86$.

\printcredits
\bibliographystyle{unsrt}
\bibliography{cas-refs}

\end{document}